\documentclass[twocolumn]{aastex6}
\usepackage{graphicx}
\usepackage{amssymb,amsfonts,amsmath,amstext,amsgen,amsopn,amsxtra,indentfirst,times}

\begin{document}

\title{Analytical Models of Exoplanetary Atmospheres. VI. Full Solutions for Improved Two-stream Radiative Transfer Including Direct Stellar Beam}

\author{Kevin Heng\altaffilmark{1}}
\author{Matej Malik\altaffilmark{1}}
\author{Daniel Kitzmann\altaffilmark{1}}
\altaffiltext{1}{University of Bern, Center for Space and Habitability, Gesellschaftsstrasse 6, CH-3012, Bern, Switzerland.  Email: kevin.heng@csh.unibe.ch}

\begin{abstract}
Two-stream radiative transfer is a workhorse in the Earth, planetary and exoplanetary sciences due to its simplicity and ease of implementation.  However, a longstanding limitation of the two-stream approximation is its inaccuracy in the presence of medium-sized or large aerosols.  This limitation was lifted by the discovery of the improved two-stream technique, where the accuracy of the scattering greenhouse effect is matched to that of multi-stream calculations by construction.  In the present study, we derive the full solutions for improved two-stream radiative transfer, following its introduction by \cite{hk17}, and include contributions from the direct stellar beam.  The generalization of the original two-stream flux solutions comes in the form of a correction factor, traditionally set to unity, which is the ratio of a pair of first Eddington coefficients.  We derive an analytical expression for this correction factor and also provide a simple fitting function for its ease of use by other workers.  We prove that the direct stellar beam is associated with a second Eddington coefficient that is on the order of unity.  Setting this second Eddington coefficient to 2/3 and $1/\sqrt{3}$ reproduces the Eddington and quadrature closures, respectively, associated with the direct beam.  We use our improved two-stream solutions for the fluxes to derive two-stream source function solutions for both the intensity and fluxes.
\end{abstract}


\section{Introduction}
\label{sect:intro}

A workhorse of radiative transfer is the two-stream method, which approximates the passage of radiation through an atmosphere as a pair of incoming and outgoing fluxes \citep{s05}.  Instead of solving a partial differential equation for the angle-dependent intensity, one solves an ordinary differential equation for the angle-integrated flux.  The price to pay is mathematical under-determination.  To mathematically close the system, one has to assume that ratios of moments of the intensity are constants known as Eddington coefficients.

There is a rich literature on two-stream radiative transfer.  \cite{joseph76} and \cite{wiscombe77} introduced ad hoc modifications to the scattering phase function to treat forward-peaked scattering of radiation due to large aerosols, known as the ``delta-Eddington approximation".  \cite{mw80} presented a unified theoretical framework that described different ways of specifying two-stream closures.  \cite{toon89} presented a specific formulation of the two-stream method that has been influential in the planetary and exoplanetary sciences (e.g., \citealt{mm99}), partly because it allows for fast computation via the inversion of a tridiagonal matrix.  \cite{toon77,toon89} introduced the ``two-stream source function" method, which reduces the radiative transfer equation from an integro-differential equation to a differential equation for the intensity by utilizing a sleight: the two-stream solution is inserted into the integral involving the scattering phase function and intensity, by assuming that the intensity is related to the two-stream flux by a constant factor.  \cite{hml14} introduced a formalism that unified the two-stream solutions with analytical solutions for temperature-pressure profiles, based partly on generalizing the relevant parts of the monograph of \cite{pierrehumbert10}.  

\begin{figure}
\begin{center}
\includegraphics[width=\columnwidth]{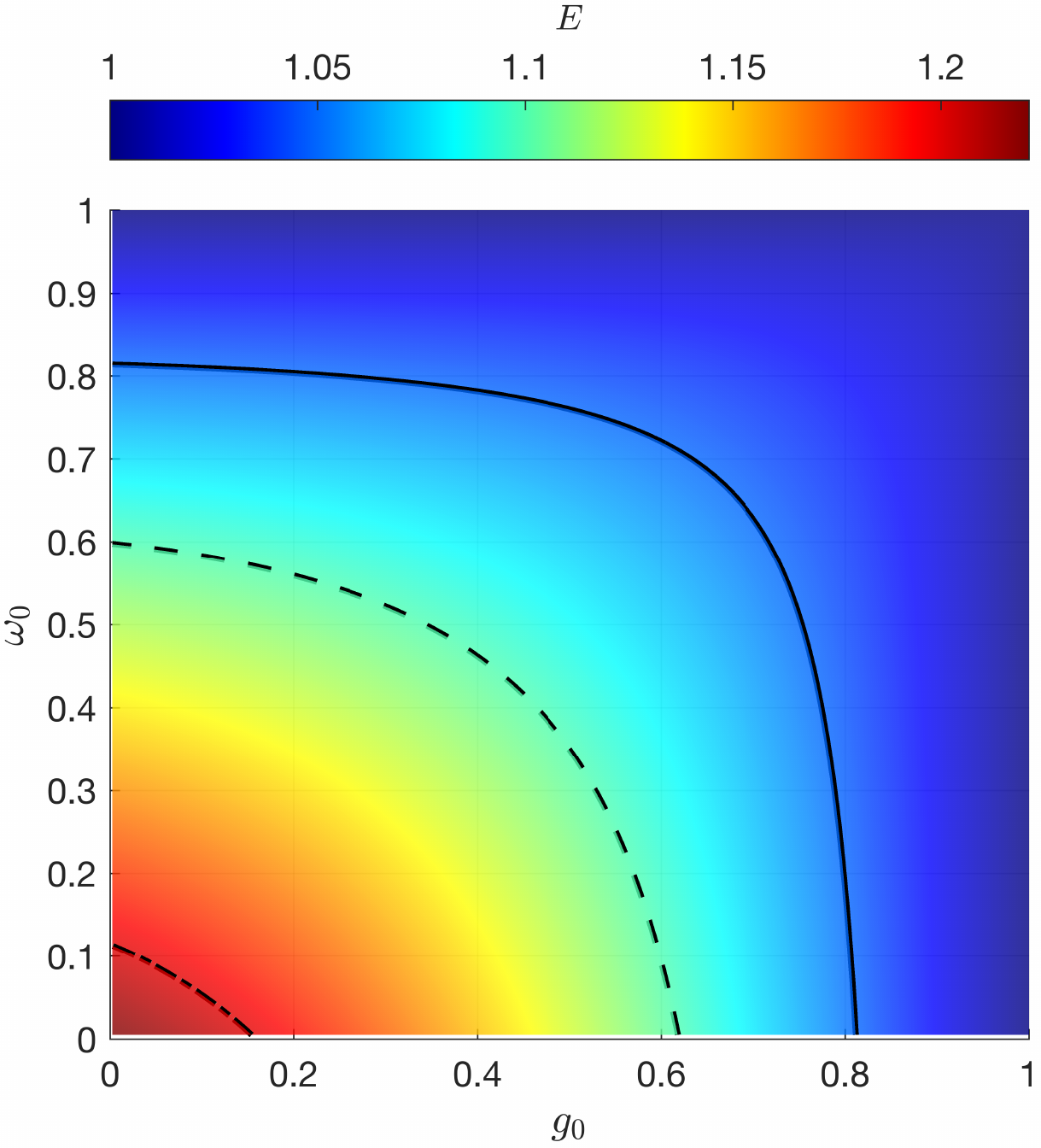}
\end{center}
\vspace{-0.1in}
\caption{Ratio of first Eddington coefficients ($E$) as a function of the single-scattering albedo ($\omega_0$) and scattering asymmetry factor ($g_0$), which is used to improve the accuracy of two-stream radiative transfer in the presence of medium-sized or large aerosols.  The solid, dashed and dot-dashed curves correspond to $E=1.05$, 1.1 and 1.2, respectively.  In the original two-stream formulation, one has $E=1$.  Pure absorption and scattering correspond to $\omega_0=0$ and 1, respectively.  Small and large aerosols are represented by $g_0=0$ and $g_0\sim 1$, respectively.  Equation (\ref{eq:efit}) provides a fitting function for $E$ that is accurate at the $\sim 0.1\%$ level.}
\vspace{0.1in}
\label{fig:efactor}
\end{figure}

\cite{kitzmann13} realized that the original two-stream method performs poorly in the presence of medium-sized or large aerosols, as it under-estimates the backscattered radiation by $\sim 10\%$.  \cite{kitzmann16} demonstrated that this shortcoming of the two-stream method translates into an over-estimation of the scattering greenhouse effect for early Mars by about 50 K.  Motivated by the work of \cite{kitzmann13} and \cite{kitzmann16}, \cite{hk17} discovered a simple improvement to the two-stream method that allows it to match the accuracy of 32-stream calculations (at the $\sim 0.01$--$1\%$ level or better, depending on the optical depth) in the presence of medium-sized or large aerosols.  

The overarching intention of the present study is to fully flesh out the improved two-stream method introduced by \cite{hk17} and also use it as input for the two-stream source function method of \cite{toon89}.  The entirety of the paper is devoted to these novel derivations.  At the heart of the improved two-stream flux solutions is a correction factor $E$, shown in Figure \ref{fig:efactor}, which is usually set to unity in the original two-stream method.  We provide a convenient fitting function in equation (\ref{eq:efit}) for the reader to compute $E$.

\section{Basic theory}

The radiative transfer equation may generally be written as \citep{chandra60,mihalas70,mihalas78,mw80,toon89}
\begin{equation}
\mu \frac{\partial I}{\partial \tau} = I - \left( 1 - \omega_0 \right) B - \omega_0 \int^{4\pi}_0 ~{\cal P} \left(I + I_{\rm beam} \right) d\Omega^\prime,
\label{eq:original}
\end{equation}
where $I$ is the wavelength-, frequency- or wavenumber-dependent intensity\footnote{The intensity is sometimes termed the ``radiance" by Earth atmospheric scientists.} (which is why we have omitted subscripts on all quantities that commit us to any of these choices, including for the Planck function), $\tau$ is the optical depth, $\omega_0$ is the single-scattering albedo and $B$ is the Planck function.  The scattering phase function, ${\cal P}(\mu^\prime, \phi^\prime, \mu, \phi)$, relates the incident polar ($\theta^\prime$) and azimuthal ($\phi^\prime$) angles to the emergent polar ($\theta$) and azimuthal ($\phi$) angles.  We further define $\mu \equiv \cos\theta$, $\mu^\prime \equiv \cos\theta^\prime$, $d\Omega \equiv d\mu ~d\phi$ and $d\Omega^\prime \equiv d\mu^\prime ~d\phi^\prime$.  Physically, one has to account for the scattered intensity coming from all incident directions.

In addition to the diffuse radiation field, we assume the presence of a direct beam of radiation due to the star, which impinges on the atmosphere with $\mu^\prime = -\mu_\star$ and $\phi^\prime = \phi_\star$.  Some portion of this incident stellar beam is absorbed, while the rest is scattered into the diffuse radiation field, which is represented by
\begin{equation}
I_{\rm beam} = F_\star \exp{\left( \frac{\tau}{\mu^\prime} \right)} ~\delta \left(\mu^\prime + \mu_\star \right) ~\delta \left(\phi^\prime - \phi_\star \right),
\end{equation}
where $F_\star$ is the incident stellar flux at the top of the atmosphere and the pair of delta functions enforce the directionality of the beam.

The radiative transfer equation is multiplied by some arbitrary function ${\cal H}(\mu)$ and integrated over all emergent angles \citep{pierrehumbert10,hml14,heng17},
\begin{equation}
\begin{split}
&\frac{\partial}{\partial \tau} \int^{2\pi}_0 \left( \int^1_0 \mu {\cal H} I ~d\mu - \int^0_{-1} \mu {\cal H} I ~d\mu \right) ~d\phi \\
&= \int^{4\pi}_0 {\cal H} I ~d\Omega - \left( 1 - \omega_0 \right) \int^{4\pi}_0 {\cal H} B ~d\Omega - {\cal I}, \\
\end{split}
\label{eq:master}
\end{equation}
where we have
\begin{equation}
\begin{split}
{\cal I} &\equiv \omega_0 \int^{4\pi}_0 \left(I + I_{\rm beam} \right) {\cal G} ~d\Omega^\prime, \\
{\cal G} &\equiv \int^{4\pi}_0 {\cal H} {\cal P} ~d\Omega.
\end{split}
\end{equation}
Equation (\ref{eq:master}) is the starting point for the suite of derivations we will perform in the current study.  The minus sign associated with the gradient terms comes about because we have rescaled the optical depth to be always positive, as has been previously elucidated in \cite{hml14} and Chapter 3.5.1 of \cite{heng17}.

Mathematically, the contribution of the direct beam to ${\cal I}$ may be evaluated in two ways: either by evaluating the delta functions first (over $d\Omega^\prime$) or by evaluating ${\cal G}$ first.  Since these mathematical operations commute, the resulting expressions must agree.  The former approach corresponds to the derivation for isotropic scattering, while the latter is for non-isotropic (or anisotropic) scattering.

\section{Improved two-stream with direct beam and isotropic scattering}

In the isotropic limit, equation (\ref{eq:original}) becomes
\begin{equation}
\mu \frac{\partial I}{\partial \tau} = I - \left( 1 - \omega_0 \right) B - \frac{\omega_0 J}{4\pi} - \omega_0 {\cal P}_\star F_\star \exp{\left( - \frac{\tau}{\mu_\star} \right)},
\end{equation}
where we define
\begin{equation}
{\cal P}_\star \equiv {\cal P} \left( -\mu_\star, \phi_\star, \mu, \phi \right).
\end{equation}
The total intensity is $J \equiv \int^{4\pi}_0 I(\mu^\prime, \tau) ~d\Omega^\prime$ and it is related to the total flux by $\epsilon = F_+/J$.  The ratio of the flux to the total intensity in one hemisphere only is denoted by $\epsilon^\prime$, where we have assumed that this ratio is the same in both hemispheres.

By integrating over the incoming ($\downarrow$) and outgoing ($\uparrow$) hemispheres, we obtain the incoming and outgoing fluxes\footnote{The flux is sometimes termed the ``irradiance" by Earth atmospheric scientists.},
\begin{equation}
\begin{split}
\frac{\partial F_\uparrow}{\partial \tau} &= \gamma_a F_\uparrow - \gamma_s F_\downarrow - \gamma_{\rm B} B - \omega_0 F_\star \chi_\uparrow \exp{\left( - \frac{\tau}{\mu_\star} \right)}, \\
\frac{\partial F_\downarrow}{\partial \tau} &= - \gamma_a F_\downarrow + \gamma_s F_\uparrow + \gamma_{\rm B} B + \omega_0 F_\star \chi_\downarrow \exp{\left( - \frac{\tau}{\mu_\star} \right)}.
\end{split}
\label{eq:pair1}
\end{equation}
The coefficients of the various terms are given by 
\begin{equation}
\begin{split}
\gamma_a &\equiv \frac{1}{\epsilon^\prime} - \frac{\omega_0}{2\epsilon} = 2E - \omega_0, \\
\gamma_s &\equiv \frac{\omega_0}{2 \epsilon} = \omega_0, \\
\gamma_{\rm B} &\equiv 2 \pi \left( 1 - \omega_0 \right).
\end{split}
\end{equation}
The derivation technique has previously been established in \cite{hml14} and also Chapter 3 of \cite{heng17}, and we will not repeat it here.  Instead, we will point out key, novel aspects of the current derivation that involve both the improvement to the two-stream method and the inclusion of the direct beam.

Previously, \cite{heng17} termed both $\epsilon$ and $\epsilon^\prime$ the ``first Eddington coefficients".  By demanding that the limiting case of an opaque, purely absorbing ($\omega_0=0$) atmosphere emits $\pi B$ of flux in each hemisphere, one obtains $\epsilon=1/2$ \citep{toon89,pierrehumbert10}.  By demanding that a purely scattering ($\omega_0=1$) atmosphere remains in radiative equilibrium, one sets $\epsilon^\prime=\epsilon$ \citep{toon89}.  However, since the two-stream solutions are formally invalid in the limit of $\omega_0=1$, the second demand is academic as has already been realized by \cite{hk17}.  Following \cite{hk17}, we set
\begin{equation}
E \equiv \frac{\epsilon}{\epsilon^\prime},
\end{equation}
and proceed to obtain an analytical expression for $E$ in terms of $\omega_0$ and the scattering asymmetry factor ($g_0$).

We have written
\begin{equation}
\begin{split}
\chi_\uparrow &\equiv \int^{2\pi}_0 \int^1_0 {\cal P}_\star ~d\mu ~d\phi, \\
\chi_\downarrow &\equiv \int^{2\pi}_0 \int^0_{-1} {\cal P}_\star ~d\mu ~d\phi,
\end{split}
\end{equation}
Since the integration involves only $\mu$ and $\phi$, we expect that ${\cal P}$ has the same mathematical behavior as ${\cal P}_\star$ with respect to this procedure.  Physically, for isotropic scattering we expect $\chi_\uparrow = \chi_\downarrow = 1/2$.  We keep these quantities distinct, because we will explicitly prove in Section \ref{sect:anisotropic} that $\chi_\uparrow = \chi_\downarrow$ is a natural outcome of setting $g_0=0$ for the more general expressions involving the gradients of the fluxes.  When scattering is non-isotropic, we generally expect that $\chi_\uparrow \ne \chi_\downarrow$, but because the scattering phase function needs to be properly normalized we have
\begin{equation}
\chi_\uparrow + \chi_\downarrow = 1.
\end{equation}
In both \cite{mw80} and \cite{toon89}, these authors have written $\gamma_3 = \chi_\uparrow$, $\gamma_4 = \chi_\downarrow$ and $\gamma_3 + \gamma_4 = 1$.

The equations governing the net ($F_- \equiv F_\uparrow - F_\downarrow$) and total ($F_+ \equiv F_\uparrow + F_\downarrow$) fluxes are
\begin{equation}
\begin{split}
\frac{\partial F_+}{\partial \tau} &= \left( \gamma_a + \gamma_s \right) F_- + \omega_0 F_\star \left( \chi_\downarrow - \chi_\uparrow \right) \exp{\left( - \frac{\tau}{\mu_\star} \right)}, \\
\frac{\partial F_-}{\partial \tau} &= \left( \gamma_a - \gamma_s \right) F_+ - 2 \gamma_{\rm B} B - \omega_0 F_\star \exp{\left( - \frac{\tau}{\mu_\star} \right)}.
\end{split}
\label{eq:pair2}
\end{equation}
We will defer seeking the solutions to $F_+$ and $F_-$, and hence $F_\uparrow$ and $F_\downarrow$, until we deal with the non-isotropic treatment, because the mathematical forms of these governing equations are almost identical.

\section{Improved two-stream with direct beam and anisotropic scattering}
\label{sect:anisotropic}

\subsection{Governing equation for net flux}

For non-isotropic scattering, we return to using equation (\ref{eq:master}) as our starting point.  If we set ${\cal H}=1$, then we obtain ${\cal G} = 1$ \citep{pierrehumbert10,hml14,heng17}.  The equation governing the net flux becomes
\begin{equation}
\frac{\partial F_-}{\partial \tau} = \left( 1 - \omega_0 \right) \left( J - 4 \pi B \right) - \omega_0 F_\star \exp{\left( - \frac{\tau}{\mu_\star} \right)}.
\end{equation}
There is an unsatisfactory ambiguity if $\epsilon \ne \epsilon^\prime$, because one could convert $J$ to the total flux using $J = F_+/\epsilon$ or $J = J_\uparrow + J_\downarrow = F_+/\epsilon^\prime$.  We resolve this ambiguity by demanding that the equation reduces to its counterpart in the isotropic limit, which yields
\begin{equation}
\frac{\partial F_-}{\partial \tau} = 2 F_+ \left( E - \omega_0 \right) - 4 \pi B \left( 1 - \omega_0 \right) - \omega_0 F_\star \exp{\left( - \frac{\tau}{\mu_\star} \right)}.
\label{eq:netflux}
\end{equation}

\subsection{Governing equation for total flux}

To obtain the equation governing the total flux, we set ${\cal H} = \mu$, which yields ${\cal G} = g_0 \mu^\prime$ \citep{pierrehumbert10,hml14,heng17}, from which it follows that
\begin{equation}
\frac{\partial K_-}{\partial \tau} = F_+ \left( 1 - \omega_0 g_0 \right) + \omega_0 g_0 \mu_\star F_\star \exp{\left( - \frac{\tau}{\mu_\star} \right)},
\end{equation}
where $K_- \equiv K_\uparrow - K_\downarrow$ and $K_\uparrow$ and $K_\downarrow$ are the second moments of the intensity.  Following \cite{hml14} and \cite{heng17}, we define a second Eddington coefficient, 
\begin{equation}
\epsilon_2 \equiv \frac{K_-}{F_+}.
\end{equation}
For the diffuse emission (which is usually dominant in the infrared range of wavelengths), we demand that $\epsilon_2 = F_+/2EF_-$ in order for the governing equation for the total flux to reduce to its counterpart in the limit of isotropic scattering.  The governing equation becomes
\begin{equation}
\frac{\partial F_+}{\partial \tau} = 2 E F_- \left( 1 - \omega_0 g_0 \right) + \gamma_\star \exp{\left( - \frac{\tau}{\mu_\star} \right)},
\label{eq:totalflux}
\end{equation}
where we have defined
\begin{equation}
\gamma_\star \equiv \frac{ \omega_0 g_0 \mu_\star F_\star}{\epsilon_2}.
\end{equation}

\subsection{Coefficients in governing equations}

For the direct beam (which is usually dominant in the optical/visible range of wavelengths), we intuitively expect that $\epsilon_2 \sim 1$.  We will now demonstrate this claim.

By comparing equations (\ref{eq:netflux}) and (\ref{eq:totalflux}) with the pair of equations in (\ref{eq:pair2}), we may infer that
\begin{equation}
\begin{split}
&\gamma_a - \gamma_s = 2 \left(E - \omega_0 \right), \\
&\gamma_a + \gamma_s = 2 E \left( 1 - \omega_0 g_0 \right),
\end{split}
\end{equation}
which yields
\begin{equation}
\begin{split}
&\gamma_a = 2E - \omega_0 \left(1 + E g_0 \right), \\
&\gamma_s = \omega_0 \left( 1 - E g_0 \right).
\end{split}
\end{equation}

In the limit of $g_0=0$ (isotropic scattering), the beam term in equation (\ref{eq:totalflux}) vanishes.  By comparing to the first equation in (\ref{eq:pair2}), we obtain $\chi_\downarrow = \chi_\uparrow$.  

Another useful comparison is between equations (\ref{eq:netflux}) and (\ref{eq:totalflux}) and the pair of equations in (\ref{eq:pair1}), which yields
\begin{equation}
\begin{split}
\chi_\uparrow &= \frac{1}{2} \left( 1 - \frac{\mu_\star g_0}{\epsilon_2} \right), \\
\chi_\downarrow &= \frac{1}{2} \left( 1 + \frac{\mu_\star g_0}{\epsilon_2} \right).
\end{split}
\end{equation}
When we compare $\chi_\uparrow$ with the $\gamma_3$ values listed in Table 1 of both \cite{mw80} and \cite{toon89}, we find that $\epsilon_2 = 2/3$ reproduces the Eddington closure, while $\epsilon_2 = 1/\sqrt{3} \approx 0.58$ reproduces the quadrature closure.  Regardless of the choice of value for the second Eddington coefficient, we always have $\chi_\uparrow + \chi_\downarrow=1$, which ensures that the scattering phase function is normalized properly.

The physical interpretation of the second Eddington coefficient is that, in the limit of $\epsilon_2/\mu_\star=1$, an extra fraction $g_0/2$ of the stellar direct beam gets scattered into the forward/incoming direction, such that when $g_0=1$ all of the stellar beam is deposited downwards.  Generally, this fraction is $\mu_\star g_0 / 2 \epsilon_2$, which allows for backscattered stellar radiation even when $g_0 = 1$ by setting $\mu_\star/\epsilon_2 \ne 1$.

\subsection{Solutions for the two-stream fluxes}

The recipe to solve for the fluxes has previously been established by both \cite{hml14} and Chapter 3 of \cite{heng17}.  We will only highlight the key points and novel generalizations.  One first merges equations (\ref{eq:netflux}) and (\ref{eq:totalflux}) into a single second-order ordinary differential equation for $F_+$.  The solution is
\begin{equation}
\begin{split}
F_+ =& A_1 \exp{\left(\alpha \tau \right)} + A_2 \exp{\left(-\alpha \tau \right)} \\
&+ 2 \pi \left( \frac{1 - \omega_0}{E - \omega_0} \right) \left[ B_i + B^\prime \left( \tau - \tau_i \right) \right] \\
&+ C_\star \exp{\left( - \frac{\tau}{\mu_\star} \right)},
\end{split}
\end{equation}
where we have $\alpha \equiv \sqrt{(\gamma_a + \gamma_s)(\gamma_a - \gamma_s)}$ and the coefficients $A_1$ and $A_2$ are determined by applying boundary conditions to the solution.  A linear expansion in the Planck function is performed to allow for the two-stream solutions to treat non-isothermal atmospheric layers described by a gradient $B^\prime \equiv \partial B/\partial \tau$, which has been shown to significantly increase computational efficiency \citep{malik17}.  The index $i=1,2$ will be explained shortly.  Note the correction factor in front of the Planck function associated with $E \ne 1$.  The coefficient $C_\star$ is derived using the method of undetermined coefficients,
\begin{equation}
C_\star = \frac{\omega_0 F_\star \left[ 2E \left( 1 - \omega_0 g_0 \right) + g_0/\epsilon_2 \right]}{4 E \left( E - \omega_0 \right) \left( 1 - \omega_0 g_0 \right) - 1/\mu_\star^2}.
\label{eq:cstar}
\end{equation}
From the expressions for $F_\uparrow$ and $F_\downarrow$, one realizes that a more compact way of expressing the algebra is to define the coupling coefficients,
\begin{equation}
\zeta_\pm \equiv \frac{1}{2} \left[ 1 \pm \sqrt{\frac{E - \omega_0}{E \left( 1 - \omega_0 g_0 \right)}} \right].
\label{eq:coupling}
\end{equation}
These are termed coupling coefficients, because they couple the outgoing and incoming fluxes in the presence of scattering (see Section 3.2 of \citealt{heng17}).  As already described in \cite{toon89}, $C_\star$ becomes indeterminate when the denominator in equation (\ref{eq:cstar}) becomes zero, which occurs when $\mu_\star = 1/2\sqrt{E (E - \omega_0)(1 - \omega_0 g_0)}$.  \cite{toon89} suggests that, ``In practice, if the equality happens to occur, this problem can be eliminated by simply choosing a slightly different value of $\mu_\star$."

The final two-stream solutions for the incoming and outgoing fluxes are expressed in the convention of a pair of atmospheric layers, subscripted by ``1" and ``2" with the former residing above the latter.  In this convention, $F_{\uparrow_2}$ and $F_{\downarrow_1}$ are the boundary conditions, which are used to eliminate $A_1$ and $A_2$.  This procedure is documented in Section 3.8.2 of \cite{heng17}.  When the dust has settled on the algebra, the two-stream solutions are
\begin{equation}
\begin{split}
&F_{\uparrow_1} = \frac{1}{\left( \zeta_- {\cal T} \right)^2 - \zeta_+^2} \left\{ \left( \zeta_-^2 - \zeta_+^2 \right) {\cal T} F_{\uparrow_2} - \zeta_- \zeta_+ \left( 1 - {\cal T}^2 \right) F_{\downarrow_1} \right. \\
&+ \left. \Pi \left[ {\cal B}_{1+} \left( \zeta^2_- {\cal T}^2 - \zeta_+^2 \right) + {\cal B}_{2+} {\cal T}  \left( \zeta_+^2 - \zeta_-^2 \right) \right. \right. \\
&+ \left. \left. {\cal B}_{1-} \zeta_- \zeta_+ \left( 1 - {\cal T}^2 \right) \right]  + C_- {\cal T} {\cal T}_\star {\cal T}_{\rm above} \left( \zeta_+^2 - \zeta_-^2 \right) \right. \\
&+ \left. {\cal T}_{\rm above} \left[ \zeta_+ \left( C_+ \zeta_- - C_- \zeta_+ \right) + \zeta_- {\cal T}^2 \left( C_- \zeta_- - C_+ \zeta_+ \right) \right] \right\},  \\
\end{split}
\label{eq:twostream_solutions_up}
\end{equation}
and
\begin{equation}
\begin{split}
&F_{\downarrow_2} = \frac{1}{\left( \zeta_- {\cal T} \right)^2 - \zeta_+^2} \left\{ \left( \zeta_-^2 - \zeta_+^2 \right) {\cal T} F_{\downarrow_1} - \zeta_- \zeta_+ \left( 1 - {\cal T}^2 \right) F_{\uparrow_2} \right. \\
&+ \left. \Pi \left[ {\cal B}_{2-} \left( \zeta^2_- {\cal T}^2 - \zeta_+^2 \right) + {\cal B}_{1-} {\cal T}  \left( \zeta_+^2 - \zeta_-^2 \right) \right. \right. \\
&+ \left. \left. {\cal B}_{2+} \zeta_- \zeta_+ \left( 1 - {\cal T}^2 \right) \right] + C_+ {\cal T}  {\cal T}_{\rm above} \left( \zeta_+^2 - \zeta_-^2 \right) \right. \\
& \left. {\cal T}_\star {\cal T}_{\rm above} \left[ \zeta_+ \left( C_- \zeta_- - C_+ \zeta_+ \right) + \zeta_- {\cal T}^2 \left( C_+ \zeta_- - C_- \zeta_+ \right) \right] \right\},  \\
\end{split}
\label{eq:twostream_solutions_down}
\end{equation}
where, for compactness of notation, we have defined
\begin{equation}
\begin{split}
\Pi &\equiv \pi \left( \frac{1 - \omega_0}{E - \omega_0} \right), \\
C_\pm &\equiv \frac{1}{2} \left[ C_\star \pm \frac{\left( C_\star/\mu_\star + \gamma_\star \right)}{2E \left( 1 - \omega_0 g_0 \right)} \right], \\
\Delta &\equiv \tau_2 - \tau_1, \\
{\cal T} &\equiv \exp{\left( -\alpha \Delta \right)}, \\
{\cal T}_\star &\equiv \exp{\left( -\frac{\Delta}{\mu_\star} \right)}, \\
{\cal T}_{\rm above} &\equiv \exp{\left(-\frac{\tau_1}{\mu_\star} \right)}, \\
{\cal B}_i &\equiv B_i \pm \frac{B^\prime}{2E \left( 1 - \omega_0 g_0 \right)}.
\end{split}
\end{equation}
The optical depth of the atmospheric layer is given by $\Delta$.  The transmissivity of the layer is represented by ${\cal T}$.  Extinction of the direct stellar beam above and through the layer are given by ${\cal T}_{\rm above}$ and ${\cal T}_\star$, respectively.  The gradient of the Planck function is $B^\prime = (B_2 - B_1)/\Delta$.

\subsection{A useful limit: $g_0=E=1$}

It is instructive to examine the limit of $g_0=1$ and $E=1$, because in the two-stream approximation it behaves as if one is in the limit of pure absorption.  In this limit, the coupling coefficients become $\zeta_-=0$ and $\zeta_+=1$, implying that several of the direct beam terms in equations (\ref{eq:twostream_solutions_up}) and (\ref{eq:twostream_solutions_down}) vanish.  For the purpose of developing physical intuition, we further let $\mu_\star = \epsilon_2$, which yields $\gamma_\star = \omega_0 F_\star$ and
\begin{equation}
C_\star = \frac{\omega_0 F_\star}{2 \left( 1 - \omega_0 \right) - 1/\epsilon_2}.
\end{equation}
Furthermore, we get $C_- = 0$ and $C_+ =C_\star$, which implies that the beam terms in the outgoing flux solution ($F_{\uparrow_1}$) vanish.  The only non-vanishing beam terms in the incoming flux solution ($F_{\downarrow_2}$) are $C_\star {\cal T}_{\rm above} ({\cal T}_\star - {\cal T})$.  One may verify that regardless of whether $2 (1 - \omega_0) > 1/\epsilon_2$ or $2 (1 - \omega_0) < 1/\epsilon_2$, $C_\star {\cal T}_{\rm above} ({\cal T}_\star - {\cal T})$ is always positive.


\section{The $E$-factor}

The ratio of the first Eddington coefficients, $E$, lies at the heart of our improved two-stream method.  In the optically-thick limit, the fraction of flux reflected by an atmospheric layer is $\zeta_-/\zeta_+$, a fact that may be verified by setting ${\cal T}=0$ in equations (\ref{eq:twostream_solutions_up}) and (\ref{eq:twostream_solutions_down}) and ignoring the blackbody and direct-beam terms.  The expression for $R_\infty = \zeta_-/\zeta_+$ may be manipulated to obtain an analytical expression for $E$.  In other words, $E$ is \textit{constructed} to reproduce the reflectivity in the optically-thick limit ($R_\infty$), as long as one has a way to numerically evaluate $R_\infty$.

Previously, \cite{hk17} performed a visual inspection of $\gamma_a = 2 - \omega_0 (1 + g_0)$ and reasoned that it generalized to $\gamma_a = 2E - \omega_0 (1 + g_0)$, while leaving $\gamma_s$ unchanged.  This led to $\zeta_\pm = \frac{1}{2} [1 \pm \sqrt{(E-\omega_0)/(E-\omega_0 g_0)} ]$ and $E = \omega_0 (1 - g_0 r^2)/(1-r^2)$.  With the formal, complete derivation performed in the present study, we obtain a different expression for $E$,
\begin{equation}
E = \frac{\omega_0}{1 - r^2 \left( 1 - \omega_0 g_0 \right)},
\label{eq:efactor}
\end{equation}
where $r \equiv (1 - R_\infty)/(1 + R_\infty)$.  

The two-stream fluxes depend on $\omega_0$ and $g_0$ only via the coupling coefficients $\zeta_-$ and $\zeta_+$.  When one inserts either expression for $E$ into its corresponding expression for $\zeta_\pm$, one obtains an identical result,
\begin{equation}
\zeta_\pm = \frac{1 \pm r}{2}.
\end{equation}
In other words, $E$ is \textit{constructed} such that the coupling coefficients only depend on $R_\infty$.  Similarly, because of the ambiguity with relating $J$, $J_\uparrow$ and $J_\downarrow$ to the fluxes via $\epsilon$ and $\epsilon^\prime$, there are multiple ways of expressing the coupling coefficients in terms of $E$.  Nevertheless, they all reduce to the same expression for $\zeta_\pm$ as stated above.  This property implies that the results shown in Figures 2, 3 and 4 of \cite{hk17} are identical\footnote{In a departure from \cite{hk17}, we do not modify the expression for ${\cal T}$ in an ad hoc way, which will produce small differences at optical depths of order unity and less.} even when the expression for $E$ in equation (\ref{eq:efactor}) is used, which is why we have not reproduced them in the current study.  The conclusions of \cite{hk17} remain qualitatively \textit{and} quantitatively unchanged.

Figure \ref{fig:efactor} shows the calculations of $E$ using equation (\ref{eq:efactor}).  $R_\infty$ is numerically evaluated using a brute-force, 32-stream discrete ordinates method \citep{s88} as implemented by the \texttt{DISORT} code \citep{hamre13}.  In these numerical calculations, one has to specify an explicit form of the scattering phase function.  A Henyey-Greenstein scattering phase function is used when $g_0 \ne 0$; when $g_0=0$, we instead use the scattering phase function associated with Rayleigh scattering.  To empower the reader in reproducing our calculations of $E$, we provide a fitting function,
\begin{equation}
\begin{split}
E =& 1.225 - 0.1582 g_0 - 0.1777 \omega_0 - 0.07465 g_0^2 \\
&+ 0.2351 \omega_0 g_0 - 0.05582 \omega_0^2.
\end{split}
\label{eq:efit}
\end{equation}
The mean and maximum errors associated with this fitting function are about 0.17\% and 1.12\%, respectively, for $g_0 \ne 0$.  For $g_0=0$, the error may be as high as 2\%.

\section{Improved two-stream source function method}

Starting from equation (\ref{eq:original}), it is always possible to rewrite the radiative transfer equation as
\begin{equation}
\begin{split}
&\frac{\partial}{\partial \tau} \left[ I ~\exp{\left( -\frac{\tau}{\mu} \right)} \right] = - \frac{\left( 1 - \omega_0 \right)}{\mu} B ~\exp{\left( -\frac{\tau}{\mu} \right)} \\
&- \frac{\omega_0}{\mu} ~\exp{\left( -\frac{\tau}{\mu} \right)} \int^{4\pi}_0 ~{\cal P} \left(I + I_{\rm beam} \right) ~d\Omega^\prime.
\end{split}
\label{eq:rt_mod}
\end{equation}
In the case of pure absorption ($\omega_0=0$), there is an exact solution to the preceding equation if one performs a series expansion of $B$ in terms of $\tau$ (or assume it to be constant).  Generally (when $\omega_0 \ne 0$), the integral is challenging to evaluate.

The trick employed by the two-stream source function method of \cite{toon89} is to argue that $I \propto F_\uparrow,F_\downarrow$ in the integral.  The scattering phase function is assumed to be $(1 + g_0)/4\pi$ and $(1 - g_0)/4\pi$ in the forward and reverse directions, respectively.  Since the two-stream fluxes do not depend on $\Omega^\prime$, the integral becomes trivial to evaluate.  The direct beam term is ignored, because the two-stream source function method was originally designed to treat diffuse thermal emission.

If we focus only on the integral term involving the diffuse emission,
\begin{equation}
\omega_0 \int^{4\pi}_0 ~{\cal P} I d\Omega^\prime = \frac{\omega_0}{2 \pi} \left[ \left( 1 \pm g_0 \right) F_\uparrow + \left( 1 \mp g_0 \right) F_\downarrow \right],
\end{equation}
where we need $I = F_{\uparrow\downarrow}/\pi$ in order for the preceding expression to reduce correctly to its isotropic limit of $\omega_0 J / 4 \pi$ when $g_0=0$.  This is a departure from \cite{toon89}, who assumed $I = F_{\uparrow\downarrow}/\epsilon$.  The minus and plus signs depend on whether one is writing the integral term for the outgoing ($\uparrow$) or incoming ($\downarrow$) flux.

We proceed somewhat differently from previous studies.  In \cite{toon77}, who first introduced the two-stream source function approximation, the form of their governing equations and the absence of $g_0$ in them suggest that their formalism does not generally treat $g_0 \ne 0$ scenarios.  In \cite{toon89}, who generalized their formalism to also consider $g_0 \ne 0$ values, their equations (53) and (54) imply that their two-stream fluxes may be written as exponentials and linear terms involving the optical depth.  The coefficients of these terms do not appear to depend on the optical depth.  This is contrary to what we find for our two-stream fluxes in equations (\ref{eq:twostream_solutions_up}) and (\ref{eq:twostream_solutions_down}), where the denominator contains the transmission function.  

In our approach, we consider the two-stream fluxes to already have undergone the integration from $\tau_1$ to $\tau_2$, such that 
\begin{equation}
\begin{split}
&\int^{\tau_2}_{\tau_1} \frac{\omega_0}{\mu} ~\exp{\left( -\frac{\tau}{\mu} \right)} \int^{4\pi}_0 ~{\cal P} I  ~d\Omega^\prime ~d\tau \\
&= \frac{\omega_0}{2 \pi \mu} \left[ \left( 1 \pm g_0 \right) F_{\uparrow_1} + \left( 1 \mp g_0 \right) F_{\downarrow_2} \right] \int^{\tau_2}_{\tau_1} \exp{\left( -\frac{\tau}{\mu} \right)} ~d\tau.
\end{split}
\end{equation}
Note that $F_{\uparrow_1}$ and $F_{\downarrow_2}$ are our two-stream fluxes given by equations (\ref{eq:twostream_solutions_up}) and (\ref{eq:twostream_solutions_down}), respectively.

If we again define $\Delta \equiv \tau_2 - \tau_1$ and
\begin{equation}
{\cal T}_0 \equiv \exp{\left( -\frac{\Delta}{\mu} \right)},
\end{equation}
then the solution to equation (\ref{eq:rt_mod}) is
\begin{equation}
\begin{split}
I_1 =& I_2 {\cal T}_0 + \left( B_i - B^\prime \tau_i \right) \left( 1 - \omega_0 \right) \left( 1 - {\cal T}_0 \right) \\
&+ B^\prime \left( 1 - \omega_0 \right) \left[ \mu \left( 1 - {\cal T}_0 \right) + \tau_1 - \tau_2 {\cal T}_0 \right] \\
&+ \frac{\omega_0}{2 \pi} \left[ \left( 1 \pm g_0 \right) F_{\uparrow_1} + \left( 1 \mp g_0 \right) F_{\downarrow_2} \right] \left( 1 - {\cal T}_0 \right),
\end{split}
\label{eq:tssf_intensity}
\end{equation}
where we have $i=1$ and 2 for the incoming and outgoing directions, respectively.  In the preceding expression, $\mu$ may be either positive or negative.

From the two-stream source function solution for the intensity in equation (\ref{eq:tssf_intensity}), one may integrate over $\mu$ to obtain the two-stream-source-function fluxes\footnote{Here, ``two-stream-source-function flux" refers to the fluxes derived using the two-stream source function approximation.},
\begin{equation}
\begin{split}
&\hat{F}_{\uparrow_1} = F_{\uparrow_2} {\cal T}^\prime + \pi B_2 \left( 1 - \omega_0 \right) \left( 1 - {\cal T}^\prime \right) \\
&+ \pi B^\prime \left( 1 - \omega_0 \right) \left\{ \frac{2}{3} \left[ 1 - \exp{\left(-\Delta \right)} \right]  - \Delta \left( 1 - \frac{{\cal T}^\prime}{3} \right) \right\} \\
&+ \frac{\omega_0}{2} \left[ \left( 1 + g_0 \right) F_{\uparrow_1} + \left( 1 - g_0 \right) F_{\downarrow_2} \right] \left( 1 - {\cal T}^\prime \right), \\
&\hat{F}_{\downarrow_2} = F_{\downarrow_1} {\cal T}^\prime + \pi B_1 \left( 1 - \omega_0 \right) \left( 1 - {\cal T}^\prime \right) \\
&- \pi B^\prime \left( 1 - \omega_0 \right) \left\{ \frac{2}{3} \left[ 1 - \exp{\left(-\Delta \right)} \right]  - \Delta \left( 1 - \frac{{\cal T}^\prime}{3} \right) \right\} \\
&+ \frac{\omega_0}{2} \left[ \left( 1 - g_0 \right) F_{\uparrow_1} + \left( 1 + g_0 \right) F_{\downarrow_2} \right] \left( 1 - {\cal T}^\prime \right), \\
\end{split}
\label{eq:tssf_fluxes}
\end{equation}
where we have used the symbols $\hat{F}_{\uparrow_1}$ and $\hat{F}_{\downarrow_2}$ to distinguish the two-stream-source-function fluxes from the two-stream fluxes.  We have defined 
\begin{equation}
{\cal T}^\prime \equiv 2 E_3\left( \Delta \right) \equiv 2 \int^\infty_1 x^{-3} \exp{\left( - x ~\Delta \right)} ~dx,
\end{equation}
where $E_3$ is the exponential integral of the third order \citep{abram,arfken}.  The reversal in the minus and plus signs for the two-stream-source-function fluxes accounts for the change in forward versus reverse directions for the incoming versus outgoing fluxes.  For example, when $g_0=1$, there should be no correction due to backscattering and all of the scattered flux should be in the forward direction.

\section{Summary and user's manual}

The computational implementation of the two-stream solutions of radiative transfer has previously been described elsewhere (e.g., \citealt{malik17}).  Instead, we focus on summarizing the different variants of the two-stream method from a practical standpoint.
\begin{itemize}

\item If the reader is interested in modeling only diffuse infrared emission in the absence of medium-sized or large aerosols, the original two-stream formulation may suffice.  For a concise statement of the two-stream fluxes with the hemispheric closure, see equation (3.58) of \cite{heng17}.  Alternatively, use equations (\ref{eq:twostream_solutions_up}) and (\ref{eq:twostream_solutions_down}) of the present study with $E=1$ and discard/ignore the direct beam terms. 

\item If the reader is interested in including the direct stellar beam in the original two-stream solutions, see \cite{mw80} and \cite{toon89}.

\item If the reader is interested in accurately modeling diffuse infrared emission in the presence of medium-sized or large aerosols, then our improved two-stream formulation in equations (\ref{eq:twostream_solutions_up}) and (\ref{eq:twostream_solutions_down}) should be used.  Consult Table 1 of \cite{mw80} or \cite{toon89} for the various choices of closure associated with the direct stellar beam (i.e., value of $\epsilon_2$).  The $E$-factor in equation (\ref{eq:efactor}) should be used in tandem with the coupling coefficients in equation (\ref{eq:coupling}).   For convenience, we provide a simple fitting function for $E$ in equation (\ref{eq:efit}).

\item Our improved two-stream solutions may be used to derive the two-stream source function solutions, which we present in equations (\ref{eq:tssf_intensity}) and (\ref{eq:tssf_fluxes}) for the intensity and fluxes, respectively.

\end{itemize}

For completeness, the summary of key symbols used and their correspondence to the symbols used in \cite{toon89} are listed in Table 1.  Appendix \ref{append:errors} lists some typographical errors we found in \cite{mw80} and \cite{toon89}.

\begin{table*}
\label{tab:symbols}
\begin{center}
\caption{Summary of symbols used, including correspondence to \cite{toon89}}
\begin{tabular}{lccc}
\hline
\hline
Symbol & Name & Symbol Used By \cite{toon89} \\
\hline
\hline
$\omega_0$ & single-scattering albedo & $\omega_0$ \\
$g_0$ & scattering asymmetry factor & $g$ \\
$\mu_\star$ & cosine of stellar beam angle & $\mu_0$ \\
$F_\star$ & stellar flux & $\pi F_s$ \\
\hline
$\epsilon=1/2$ & first Eddington coefficient & $\mu_1$ \\
$\epsilon^\prime$ & first Eddington coefficient & --- \\
$\epsilon_2 \sim 1$ & second Eddington coefficient & --- \\
$E \equiv \epsilon/\epsilon^\prime$ & ratio of first Eddington coefficients & 1 \\
$\zeta_\pm$ & coupling coefficients & --- \\
$\zeta_-/\zeta_+$ & asymptotic value of reflectivity & $\Gamma$ \\
$A_1 \zeta_+$ & --- & $k_1$ \\
$A_2 \zeta_-$ & --- & $k_2 \Gamma$ \\
\hline
$\Delta \equiv \tau_2 - \tau_1$ & optical depth of atmospheric layer & $\tau$ \\
${\cal T} \equiv \exp{\left(-\alpha \Delta\right)}$ & transmissivity or transmission function & $\exp{\left(-\lambda \tau \right)}$ \\
${\cal T}_\star \equiv \exp{\left(-\Delta/\mu_\star \right)}$ & extinction of direct beam through atmospheric layer & $\exp{\left(-\tau/\mu_0 \right)}$ \\
${\cal T}_{\rm above}$ & extinction of direct beam above atmospheric layer & $\exp{\left(-\tau_c/\mu_0 \right)}$ \\
$F_\uparrow$ & outgoing flux & $F^+$ \\
$F_\downarrow$ & incoming flux & $F^-$ \\
$F_- \equiv F_\uparrow - F_\downarrow$ & net flux & $F_{\rm net}$ \\
$F_+ \equiv F_\uparrow + F_\downarrow$ & total flux & --- \\
\hline
$\gamma_a$ & coefficient in equation for $\frac{\partial F_{\uparrow \downarrow}}{\partial \tau}$ & $\gamma_1$ \\
$\gamma_s$ & coefficient in equation for $\frac{\partial F_{\uparrow \downarrow}}{\partial \tau}$ & $\gamma_2$ \\
$\gamma_{\rm B}$ & coefficient in equation for $\frac{\partial F_{\uparrow \downarrow}}{\partial \tau}$ & $2\pi \left( 1 - \omega_0 \right)$ \\
$\gamma_\star \equiv \omega_0 g_0 \mu_\star F_\star / \epsilon_2$ & coefficient of direct beam term in equation for $\frac{\partial F_+}{\partial \tau}$ & --- \\
$\alpha \equiv \sqrt{\left(\gamma_a+\gamma_s\right)\left(\gamma_a-\gamma_s\right)}$ & diffusivity factor & $\lambda$ \\
$\chi_\uparrow$ & --- & $\gamma_3 = \beta_{0\nu}$ \\
$\chi_\downarrow$ & --- & $\gamma_4 = 1 -  \beta_{0\nu}$ \\
\hline
\hline
\end{tabular}\\
\end{center}
\end{table*}

\begin{acknowledgments}

We acknowledge financial support from the Swiss National Science Foundation, the European Research Council via Consolidator Grant ERC-2017-CoG-771620-EXOKLEIN (awarded to KH), the PlanetS National Center of Competence in Research (NCCR), the Center for Space and Habitability (CSH) and the Swiss-based MERAC Foundation.

\end{acknowledgments}

\appendix

\section{Typographical errors in Meador \& Weaver (1980) and Toon et al. (1989)}
\label{append:errors}

\subsection{\cite{toon89}}

We use the same notation as \cite{toon89}: $(\mu^\prime,\phi^\prime)$ and $(\mu,\phi)$ for the incident and emergent directions, respectively.  Equation (1) of \cite{toon89} concisely states the functional dependence of their scattering phase function: $P_\nu(\mu, \mu^\prime, \phi, \phi^\prime)$.  The scattering phase function associated with the direct beam is stated in their equation (3): $P_\nu(\mu, -\mu_0, \phi, \phi_0)$.  Based on their stated convention and also their equation (10), we may conclude that equation (9) of \cite{toon89} is in error as it should instead be
\begin{equation}
\beta_{0\nu} = \frac{1}{2} \int^1_0 ~P_\nu\left(\mu, -\mu_0\right) ~d\mu.
\end{equation}
Specifically, the integration should be over $\mu$ and not $\mu^\prime$.

In equation (12), there is a typographical error: $S^+_n$ should instead be $S^-_n$.  The logical flow of the text into equations (13) and (14) of \cite{toon89} makes this clear.

\subsection{\cite{mw80}}

Our notation for $\mu$, $\mu^\prime$, $\phi$ and $\phi^\prime$ are the same as that of \cite{mw80}.  Equation (1) of \cite{mw80} concisely states the functional dependence of their scattering phase function: $P_\nu(\mu, \phi; \mu^\prime, \phi^\prime)$.  It also states the functional dependence of the scattering phase function associated with the direct beam: $P_\nu(\mu, \phi; -\mu_0, \phi_0)$.  Equation (7) of \cite{mw80} follows logically from their equation (3).  However, their equation (9) suffers from the same logical inconsistency as what we pointed out for \cite{toon89}, which is that the integration should be over $\mu$ and not $\mu^\prime$.  This is clearly seen in the transition from equation (7) to equation (10).  It is not an issue of a mere change of notation for the integration variable, because it needs to correspond to the correct variable in the scattering phase function.  Equation (9) of \cite{mw80} should instead be
\begin{equation}
\beta_ i= \frac{1}{2 \omega_0} \int^1_0 ~P_\nu\left(\mu, -\mu_i \right) ~d\mu.
\end{equation}
It is likely that this error in \cite{mw80} propagated into \cite{toon89}.

\label{lastpage}

\end{document}